\documentstyle[prc,aps,preprint,epsfig]{revtex}
\tighten
\begin{document}
\draft
\title{A third family of super dense stars in the presence of antikaon 
condensates}
\author{Sarmistha Banik and Debades Bandyopadhyay}
\address{Saha Institute of Nuclear Physics, 1/AF Bidhannagar, 
Calcutta 700 064, India}

\maketitle

\begin{abstract}
The formation of $K^-$ and $\bar K^0$ condensation in $\beta$-equilibrated
hyperonic matter is investigated within a relativistic mean field model. In
this model, baryon-baryon and (anti)kaon-baryon interactions are mediated by
the exchange of mesons. It is found that antikaon condensation is not only
sensitive to the equation of state but also to antikaon optical potential 
depth. For large values of antikaon optical potential depth, $K^-$ condensation
sets in before the appearance of negatively charged hyperons. We treat $K^-$
condensation as a first order phase transition. The Gibbs criteria and
global charge conservation laws are used to describe the mixed phase. Nucleons
and $\Lambda$ hyperons behave dynamically in the mixed phase. A second order
phase transition to $\bar K^0$ condensation occurs in the pure $K^-$ condensed
phase. Along with $K^-$ condensation, $\bar K^0$ condensation makes the 
equation of state softer thus resulting in smaller maximum mass stars compared
with the case without any condensate. This equation of state also leads to a
stable sequence of compact stars called the third family branch, beyond the 
neutron star branch. The compact stars in the third family branch have 
different compositions and smaller radii than that of the neutron star branch.

\end{abstract}
\section{Introduction}
There is a growing interest to understand whether a stable sequence of 
compact stars could exist in nature beyond a neutron star branch.
>From observations, two families of compact stars - white dwarfs and neutron 
stars, are known to us. The physical reasons for stability in white dwarfs 
and neutron stars are different. In white dwarfs, it is the Fermi pressure 
of degenerate electrons that stabilizes the stars. After the white dwarf 
branch, there exists no stable star having central densities in the range 
$10^9- 10^{14} gm/ cm^3$ as is evident from the mass-radius relationship 
obtained by the Tolman-Oppenheimer-Volkoff (TOV) equations \cite{Gle}. 
The stability is regained in the neutron star branch. 
In this case, the Fermi pressure of 
degenerate and interacting baryons is responsible for the stability of 
neutron stars. The question is what is next to a neutron star branch - again
an unstable region followed by a stable configurations of super dense stars!
If this picture is true, what could be the mechanism behind the stability of 
a possible third family of compact stars? Gerlach \cite{Ger} first pointed out 
that a third 
family of compact stars could be a possibility. He noted that a jump in the 
equation of state (EoS) or equivalently a large discontinuity in the speed of 
sound beyond the central density corresponding to the maximum neutron star mass
might result in a stable family of super dense stars beyond the neutron star 
branch. However, Wheeler et al. showed that there was no stable compact stars 
beyond a neutron star branch for a smooth EoS \cite{Whe}.
  
Neutron stars are very useful laboratories to investigate the properties of 
dense matter \cite{Gle,Web}. At the core of a neutron star, the matter density 
could exceed 
by a few times normal nuclear matter density. Various exotic forms of matter 
such as hyperonic matter, quark-hadron mixed phase and Bose-Einstein 
condensation of strange particles may appear there. Earlier it was found that 
there might be kinks in the EoS because of the appearance of an exotic form of 
matter
in the high density regime. Consequently there is a discontinuity in the speed 
of sound. This motivated various groups to explore whether physical processes 
like a first order phase transition from hadronic matter to quark matter or 
hadronic matter to hyperonic matter could give rise to an EoS leading to a 
third family of compact stars \cite{Gle00,Sche,Han,Fra}. It was shown by 
various authors that a first 
order quark-hadron phase transition could produce an EoS such that the neutron 
star branch is terminated by the softening in the EoS due to the mixed phase 
and the third family of compact stars attained a pure quark phase in the 
core \cite{Gle00,Sche,Fra}.
Similarly, a first order phase transition from hadronic matter to hyperonic 
matter
led to stable solutions beyond the neutron star branch for certain parameters
sets of hyperon-hyperon interactions \cite{Han}. It was demonstrated in those 
calculations 
that non-identical stars of same mass but distinctly different radii and 
compositions 
could exist because of partial overlapping mass regions of the neutron star 
branch 
and the third family branch. These pairs are known as "neutron star twins"
\cite{Gle00,Sche}.

Besides quark-hadron phase transition and the formation of hyperonic matter, 
Bose-Einstein condensation of antikaons may occur in the $\beta$-stable and 
charge neutral dense matter. This antikaon condensation could be a first or 
second order phase transition \cite{Gle99}. 
It was first demonstrated by Kaplan and Nelson \cite{Kap} using a chiral 
$SU(3)_L\times SU(3)_R$ Lagrangian that $K^-$meson may undergo 
Bose-Einstein condensation in dense hadronic matter formed in heavy ion 
collisions. The strongly attractive 
$K^-$-nucleon interaction lowers the effective mass of $K^-$ mesons in dense 
matter. Consequently, the in-medium energy of $K^-$ mesons decreases and s-wave 
condensation sets in when the in-medium energy of ${K^-}$ mesons equals to its 
chemical potential. Later, this chiral model was applied to the study of 
${K^-}$ condensation in the core of neutron stars \cite{Bro92,Thor,Ell,Pra97}. 
Also, the traditional meson exchange picture where baryons and (anti)kaons 
interact through meson exchange, was extensively exploited to investigate 
antikaon condensation in dense matter relevant to (proto)neutron
stars \cite{Gle99,Mut,Kno,Sch,Pon,Pal,Bani}. 
The onset of ${K^-}$ condensation in neutron star matter depends on the nuclear 
equation of state and also on the depth of antikaon optical potential.  

The threshold density of ${K^-}$ condensation in nucleons-only star matter is 
3-4 times normal nuclear matter density. The net effect of $K^-$ mesons in the 
condensate in neutron star matter is to maintain charge neutrality replacing 
electrons and soften the EoS resulting in the reduction of maximum
mass of the star \cite{Gle99,Thor,Mut,Kno,Sch,Pal,Bani}. In the presence of
hyperons, ${K^-}$ condensation was delayed to higher density. Recently, we have 
investigated the formation of $\bar K^0$ condensation along with ${K^-}$ 
condensation in dense nuclear and hyperonic matter relevant to (proto)neutron 
stars within Relativistic Mean Field (RMF) models \cite{Pal,Bani}. It was found 
that the onset of $\bar K^0$ condensation always occurred later than that of 
${K^-}$ condensation. With the onset of $\bar K^0$ condensation, abundances of 
neutrons and protons become equal in the high density matter \cite{Pal,Bani}. 
In the presence of hyperons, the thresholds of ${K^-}$ and $\bar K^0$ 
condensation  are shifted to higher densities. Along with ${K^-}$ condensate, 
$\bar K^0$ condensate makes the EoS much softer resulting 
in smaller maximum mass star compared to the case without any condensate.

In this paper we investigate the effects of antikaon condensation on the 
equation of state within a RMF model where baryons and (anti)kaons are 
interacting via meson exchanges. Also, we study the properties of the third 
family of compact stars with this EoS. The paper is organised in the following 
way. In section 2, we describe the RMF model of strong interactions and 
different phases of dense matter. In section 3, the parameters of the model 
are discussed. Results of our calculation and their relations to the third 
family of compact stars are explained in section 4. Section 5 deals with 
summary and conclusions. 

\section{Formalism}
In this calculation, we describe a first order phase transition from hadronic 
matter to the antikaon condensed phase in compact stars.
We adopt a relativistic field theoretical model to describe the pure hadronic
matter, pure antikaon condensed matter and the mixed phase. The constituents 
of matter - ${n,p,\Lambda,\Sigma^+,\Sigma^-,\Sigma^0, \Xi^-,\Xi^0 }$ of the
baryon octet and electrons and muons, in the hadronic phase 
are to satisfy beta-equilibrium and local charge neutrality 
condition. The strong interaction between baryons is mediated by the exchange
of scalar and vector mesons.  The model is also extended to include hyperon-
hyperon interaction through two additional hidden-strangeness mesons-
scalar meson
$f_0$(975) (denoted hereafter as $\sigma^*$) and the vector meson $\phi$(1020) 
\cite{Sch}.
Therefore the Lagrangian density for the pure hadronic phase is given by
\begin{eqnarray}
{\cal L}_B &=& \sum_B \bar\psi_{B}\left(i\gamma_\mu{\partial^\mu} - m_B
+ g_{\sigma B} \sigma - g_{\omega B} \gamma_\mu \omega^\mu
- g_{\rho B}
\gamma_\mu{\mbox{\boldmath t}}_B \cdot
{\mbox{\boldmath $\rho$}}^\mu \right)\psi_B\nonumber\\
&& + \frac{1}{2}\left( \partial_\mu \sigma\partial^\mu \sigma
- m_\sigma^2 \sigma^2\right) - U(\sigma) \nonumber\\
&& -\frac{1}{4} \omega_{\mu\nu}\omega^{\mu\nu}
+\frac{1}{2}m_\omega^2 \omega_\mu \omega^\mu
- \frac{1}{4}{\mbox {\boldmath $\rho$}}_{\mu\nu} \cdot
{\mbox {\boldmath $\rho$}}^{\mu\nu}
+ \frac{1}{2}m_\rho^2 {\mbox {\boldmath $\rho$}}_\mu \cdot
{\mbox {\boldmath $\rho$}}^\mu  + {\cal L}_{YY}~.
\end{eqnarray}

Here $\psi_B$ denotes the Dirac spinor for baryon B with vacuum mass $m_B$
and isospin operator ${\mbox {\boldmath t}}_B$. The scalar
self-interaction term \cite{Bog} is
\begin{equation}
U(\sigma) = \frac{1}{3} g_2 \sigma^3 + \frac{1}{4} g_3 \sigma^4 ~.
\end{equation}
The Lagrangian density (${\cal L}_{YY}$)
responsible for hyperon-hyperon interaction is given by,
\begin{eqnarray}
{\cal L}_{YY} &=& \sum_B \bar\psi_{B}\left(
g_{\sigma^* B} \sigma^* - g_{\phi B} \gamma_\mu \phi^\mu
\right)\psi_B\nonumber\\
&& + \frac{1}{2}\left( \partial_\mu \sigma^*\partial^\mu \sigma^*
- m_{\sigma^*}^2 \sigma^{*2}\right)
-\frac{1}{4} \phi_{\mu\nu}\phi^{\mu\nu}
+\frac{1}{2}m_\phi^2 \phi_\mu \phi^\mu~.
\label{Lag}
\end{eqnarray}

At the interior of compact stars, 
the generalised $\beta$-decay processes may be written in the form
$B_1 \longrightarrow B_2 + l+ \bar \nu_l$ and 
$B_2 +l \longrightarrow B_1 +\nu_l$ where $B_1$ and $B_2$ are baryons and 
l is a lepton. 
These weak processes conserve
baryon number and electric charge and are in chemical equilibrium.
Therefore the generic equation for chemical equilibrium condition is
\begin{equation}
\mu_i = b_i \mu_n - q_i \mu_e ~,
\end{equation}
where $\mu_n$, $\mu_e$ and $\mu_i$ are respectively
the chemical potentials of neutrons, electrons and i-th baryon with
$\mu_{i} = (k^2_{F_{i}} + m_i^{* 2} )^{1/2} + g_{\omega i} \omega_0
+ g_{\phi i} \phi_0 + I_{3i} g_{\rho i} \rho_{03}$ and $b_i$ and $q_i$ are
baryon and electric charge of ith baryon respectively.
The above equation implies that there are two independent chemical
potentials $\mu_n$ and $\mu_e$ corresponding to two conserved 
charges i.e. baryon number and electric charge.
In neutron stars, electrons are converted to muons by
$e^- \to \mu^- + \bar \nu_\mu + \nu_e$ when the electron chemical potential
becomes equal to the muon mass. Therefore, we have $\mu_e=\mu_{\mu}$ in a
neutron star. In the pure hadronic phase, the total charge density is
\begin{equation}
Q^h = \sum_B q_B n^h_B -n_e -n_\mu =0
\end{equation}
where $n_B^h$ is the number density of baryon B in the pure hadronic phase and
$n_e$ and $ n_\mu$ are charge densities of electrons and muons respectively.

Solving the equations of motion in the mean field approximation 
\cite{Ser} along with effective baryon masses ($m_i^*$) and equilibrium 
conditions 
we immediately compute the equation of state in the pure hadronic phase.
The energy density ($\varepsilon^h$) is related to the pressure
($P^h$) in this phase through the Gibbs-Duhem relation 
\begin{equation}
P^h=\sum_i \mu_i n_i -\varepsilon^h~.
\end{equation}
Here $\mu_i$ and $n_i$ are chemical potential and number density for 
i-th species.

The pure antikaon condensed phase is composed of baryons, leptons and antikaons.
In this phase, the constituents of matter are in chemical equilibrium 
under weak interactions and maintain local charge neutrality. The baryon-baryon
interaction in the pure condensed phase is described by the Lagrangian 
density as given by Eq. (1). Earlier it was shown that nucleons in the pure 
nuclear and antikaon condensed matter behaved differently because of their 
dynamical nature \cite{Gle99}. It was attributed to different mean fields which
nucleons experienced in the pure phases. We adopt a relativistic field
theoretical approach for the description of (anti)kaon-baryon interaction
\cite{Gle99}. Here, baryon-baryon and (anti)kaon-baryon interactions are 
treated on the same footing. We also include $\bar K^0$ condensation in this 
calculation and it occurs as a second order phase transition in the pure $K^-$
condensed phase. The Lagrangian density for (anti)kaons
in the minimal coupling scheme is, 
\begin{equation}
{\cal L}_K = D^*_\mu{\bar K} D^\mu K - m_K^{* 2} {\bar K} K ~,
\end{equation}
where the covariant derivative
$D_\mu = \partial_\mu + ig_{\omega K}{\omega_\mu} + ig_{\phi K}{\phi_\mu}
+ i g_{\rho K}
{\mbox{\boldmath t}}_K \cdot {\mbox{\boldmath $\rho$}}_\mu$.
The isospin doublet for kaons
is denoted by $K\equiv (K^+, K^0)$ and that for antikaons is
$\bar K\equiv (K^-, \bar K^0)$. The effective mass of (anti)kaons in this
minimal coupling scheme is given by
\begin{equation}
m_K^* = m_K - g_{\sigma K} \sigma - g_{\sigma^* K} \sigma^* ~,
\end{equation}
where $m_K$ is the bare kaon mass. In the mean field approximation 
\cite{Ser} adopted here, the meson fields are replaced by their
expectation values.
The mean meson fields are denoted by $\sigma$, $\sigma^*$, $\omega_0$,
$\phi_0$ and $\rho_{03}$.
The dispersion relation representing the in-medium energies of
$\bar K\equiv (K^-, \bar K^0)$ for $s$-wave (${\bf k}=0$) condensation
is given by
\begin{equation}
\omega_{K^-,\: \bar K^0} = m_K^* - g_{\omega K} \omega_0 - g_{\phi K} \phi_0
\mp \frac{1}{2} g_{\rho K} \rho_{03} ~,
\end{equation}
where the isospin projection $I_{3\bar K} =\mp 1/2$ for the mesons
$K^-$ ($-$ sign) and $\bar K^0$ (+ sign) are explicitly written in the
expression. Since the $\sigma$ and $\omega$ fields generally increase with
density  and both the terms containing $\sigma$ and $\omega$ fields in Eq.(9)
are attractive for antikaons, the in-medium energies of $\bar K$ mesons 
decrease in nuclear medium. On the other hand, in nucleon-only matter
$\rho_{03} \equiv n_p - n_n$ ($n_p$ and $n_n$ are the proton and
neutron densities) is negative; thus the $\rho$-meson field favors the
formation of $\bar K^0$ condensation over that of $K^-$ condensation. In
hyperonic matter, the repulsive $\phi$ meson term may delay the onsets of
antikaon condensation \cite{Sch}. 
The in-medium energies of kaons $K\equiv (K^+, K^0)$ are given by,
\begin{equation}
\omega_{K^+,\: K^0} = m_K^* + g_{\omega K} \omega_0 + g_{\phi K} \phi_0
\pm \frac{1}{2} g_{\rho K} \rho_{03} ~.
\end{equation}
It is evident here that kaon condensation may be impossible in 
$\beta$-equilibrated matter because the $\omega$-meson term is repulsive for 
kaons and dominates over the attractive $\sigma$-meson term at higher densities.
However, the attractive $\phi$ meson term may decrease kaon energies
in the presence of hyperons \cite{Sch}.

The meson field equations in the pure condensed phase 
are derived from Eqs. (1)-(3) and (7) as
\begin{eqnarray}
m_\sigma^2\sigma &=& -\frac{\partial U}{\partial\sigma}
+ \sum_B g_{\sigma B} n_B^{{\bar K},S}
+ g_{\sigma K} \sum_{\bar K} n_{\bar K} ~,\\
{m_\sigma^*}^2\sigma^* &=& \sum_B g_{\sigma^* B} n_B^{{\bar K},S}
+ g_{\sigma^* K} \sum_{\bar K} n_{\bar K} ~,\\
m_\omega^2\omega_0 &=& \sum_B g_{\omega B} n_B^{\bar K}
- g_{\omega K} \sum_{\bar K} n_{\bar K} ~,\\
m_\phi^2\phi_0 &=& \sum_B g_{\phi B} n_B^{\bar K}
- g_{\phi K} \sum_{\bar K} n_{\bar K} ~,\\
m_\rho^2\rho_{03} &=& \sum_B g_{\rho B} I_{3B} n_B^{\bar K}
+ g_{\rho K} \sum_{\bar K} I_{3\bar K} n_{\bar K} ~.
\end{eqnarray}
Here the scalar and number density of baryon $B$ in the antikaon condensed 
phase are respectively
\begin{eqnarray}
n_B^{{\bar K},S} &=& \frac{2J_B+1}{2\pi^2} \int_0^{k_{F_B}}
\frac{m_B^*}{(k^2 + m_B^{* 2})^{1/2}} k^2 \ dk ~,\\
n_B^{\bar K} &=& (2J_B+1)\frac{k^3_{F_B}}{6\pi^2} ~,
\end{eqnarray}
with effective baryonic mass $m_B^*=m_B - g_{\sigma B}\sigma
- g_{\sigma^* B}\sigma^*$,
Fermi momentum $k_{F_B}$, spin $J_B$, and isospin projection
$I_{3B}$. Note that for $s$-wave ${\bar K}$ condensation, the scalar and
vector densities of antikaons are same and those are given 
by \cite{Gle99}
\begin{equation}
n_{K^-,\: \bar K^0} = 2\left( \omega_{K^-, \bar K^0} + g_{\omega K} \omega_0 
+ g_{\phi K} \phi_0 \pm \frac{1}{2} g_{\rho K} \rho_{03} \right) {\bar K} K  
= 2m^*_K {\bar K} K  ~.
\end{equation}
The total energy density in the antikaon condensed matter
has contributions from baryons, leptons and antikaons and is given by,
\begin{eqnarray}
{\varepsilon^{\bar K}}  &=& \frac{1}{2}m_\sigma^2 \sigma^2
+ \frac{1}{3} g_2 \sigma^3
+ \frac{1}{4} g_3 \sigma^4  + \frac{1}{2}m_{\sigma^*}^2 \sigma^{*2}
+ \frac{1}{2} m_\omega^2 \omega_0^2 + \frac{1}{2} m_\phi^2 \phi_0^2
+ \frac{1}{2} m_\rho^2 \rho_{03}^2  \nonumber \\
&& + \sum_B \frac{2J_B+1}{2\pi^2}
\int_0^{k_{F_B}} (k^2+m^{* 2}_B)^{1/2} k^2 \ dk
+ \sum_l \frac{1}{\pi^2} \int_0^{K_{F_l}} (k^2+m^2_l)^{1/2} k^2 \ dk
+ \varepsilon_3  ,
\end{eqnarray}
where $l$ goes over electrons and muons and 
the energy density for antikaons is
\begin{equation}
\varepsilon_3 = m^*_K \left( n_{K^-} + n_{\bar K^0} \right) .
\end{equation}
Since antikaons form $s$-wave Bose condensates, they do not directly
contribute to the pressure so that the pressure is due to baryons and
leptons only
\begin{eqnarray}
P^{\bar K} &=& - \frac{1}{2}m_\sigma^2 \sigma^2 - \frac{1}{3} g_2 \sigma^3
- \frac{1}{4} g_3 \sigma^4  - \frac{1}{2}m_{\sigma^*}^2 \sigma^{*2}
+ \frac{1}{2} m_\omega^2 \omega_0^2 + \frac{1}{2} m_\phi^2 \phi_0^2
+ \frac{1}{2} m_\rho^2 \rho_{03}^2 \nonumber \\
&& + \frac{1}{3}\sum_B \frac{2J_B+1}{2\pi^2}
\int_0^{k_{F_B}} \frac{k^4 \ dk}{(k^2+m^{* 2}_B)^{1/2}}
+ \frac{1}{3} \sum_l \frac{1}{\pi^2}
\int_0^{K_{F_l}} \frac{k^4 \ dk}{(k^2+m^2_l)^{1/2}}
 ~.
\end{eqnarray}
The effect of antikaons in the pressure term is through the meson fields,
which change due to the presence of additional antikaon source terms in the 
equations of motions in Eqs. (11)-(15).
In the absence of those source terms for antikaons, we retain the 
equations of motion  for meson fields in the pure hadronic phase. 

With the onset of $\bar K$ condensation,
other strangeness changing processes may occur such as,
$N \rightleftharpoons N + \bar K$ and $e^- \rightleftharpoons K^- + \nu_e$,
where $N\equiv (n,p)$ and $\bar K \equiv (K^-, \bar K^0)$ denote the
isospin doublets for nucleons and antikaons, respectively. The
requirement of chemical equilibrium yields
\begin{eqnarray}
\mu_n - \mu_p &=& \mu_{K^-} = \mu_e ~, \\
\mu_{\bar K^0} &=& 0 ~,
\end{eqnarray}
where $\mu_{K^-}$ and $\mu_{\bar K^0}$ are respectively the chemical
potentials of $K^-$ and $\bar K^0$. The above conditions determine the
onsets of antikaon condensations. When the effective
energy of $K^-$ meson ($\omega_{K^-}$)
equals to its chemical potential ($\mu_{K^-}$) which, in
turn, is equal to $\mu_e$, a $K^-$ condensate
is formed. Similarly, $\bar K^0$ condensation is formed when its in-medium
energy satisfies the condition $\omega_{\bar K^0} = \mu_{\bar K^0} = 0$.

The total charge density in the antikaon condensed phase is 
\begin{equation}
Q^{\bar K}=\sum_B q_B n_B^{\bar K} -n_K^--n_e -n_\mu =0
\end{equation}

Now we describe the mixed phase of hadronic matter and $K^-$ 
condensed matter. In earlier works, the Maxwell construction was 
employed for a first order phase transition to antikaon condensation
\cite{Thor,Kno,Pal,Bani}. It was  argued that 
the Maxwell construction was adequate for a system with only one 
conserved charge \cite{Gle99,Gle92}.
However, neutron star matter has more than one conserved charge.
As evident from the generic equation of $\beta$-equilibrium (Eq. (4)) there 
are two independent chemical potentials namely $\mu_n$
and $\mu_e$ connected to the conservation of baryon number and 
electric charge, respectively. Therefore, a Maxwell construction 
in this case will result in a discontinuity in one of the chemical 
potentials \cite{Gle99,Gle92}. It was shown by Glendenning \cite{Gle99,Gle92} 
that the Gibbs conditions
for thermodynamic equilibrium along with global conservation laws 
would be required  to describe the mixed phase. For compact star matter
the Gibbs phase rules read,
\begin{eqnarray}
P^h&=& P^{\bar K},\\
\mu_B^h& =& \mu_B^{\bar K}.
\end{eqnarray}
where $\mu_B^h$ and $\mu_B^{\bar K}$ are chemical potentials of baryon B in the
pure hadronic and $K^-$ condensed phase, respectively.
The conditions of global charge neutrality and baryon number conservation are 
imposed through the relations
\begin{equation}
(1-\chi) Q^h + \chi Q^{\bar K} = 0,\\
\end{equation}
\begin{equation}
n_B=(1-\chi) n_B^h + \chi n_B^{\bar K}~,
\end{equation}
where $\chi$ is the volume fraction of $K^-$ condensed phase in the mixed 
phase. The total energy density in the mixed phase is
\begin{equation}
\epsilon=(1-\chi)\epsilon^h + \chi \epsilon^{\bar K}~.
\end{equation}

\section{Parameters}
In the effective field theoretic approach discussed here, knowledge
of three distinct sets of coupling constants for nucleons, kaons and hyperons 
associated with the exchange of scalar $\sigma$, isoscalar-vector $\omega$, and
vector-isovector $\rho$ mesons and
two additional hidden-strangeness mesons, scalar meson 
($\sigma^*$) and vector meson $\phi$ are required.
The nucleon-meson coupling constants generated by reproducing the
nuclear matter saturation properties are taken from Glendenning and
Moszkowski of Ref. \cite{Gle91}. This set is referred to as GM1 
and its parameters are also listed 
in Table I.

The vector coupling constants for hyperons are determined from SU(6)
symmetry as,
\begin{eqnarray}
\frac{1}{2}g_{\omega \Lambda} = \frac{1}{2}g_{\omega \Sigma} = g_{\omega \Xi} =
\frac{1}{3} g_{\omega N},\nonumber\\
\frac{1}{2}g_{\rho \Sigma} = g_{\rho \Xi} = g_{\rho N}{\rm ;}~~~
g_{\rho \Lambda} = 0, \nonumber\\
2 g_{\phi \Lambda} = 2 g_{\phi \Sigma} = g_{\phi \Xi} =
-\frac{2\sqrt{2}}{3} g_{\omega N}. ~
\end{eqnarray}
The scalar meson ($\sigma$) coupling to hyperons is obtained from the
potential depth of a hyperon (Y) in the saturated nuclear matter
\begin{equation}
U_Y^N(n_0) = - g_{\sigma Y} {\sigma} + g_{\omega Y} {\omega_0}.
\end{equation}
The analysis of energy levels in $\Lambda$-hypernuclei suggests a well depth of
$\Lambda$ in symmetric nuclear matter $U_{\Lambda}^N(n_0)=-30$ MeV 
\cite{Chr,Dov}. On the other hand, recent analysis of a few $\Xi$-hypernuclei 
events in emulsion experiments predicts a $\Xi$ well depth of
$U_{\Xi}^N(n_0)=-18$ MeV \cite{Fak,Kha} in normal nuclear matter.
However, the situation for the $\Sigma$ potential depth in normal nuclear 
matter is very unclear.  The only known bound $\Sigma$-hypernucleus is 
$^4_{\Sigma}He$ \cite{Hay}. The most updated analysis of $\Sigma^-$ atomic data 
indicates a strong isoscalar repulsion in $\Sigma$-nuclear matter interaction 
\cite{Fri94}.  Therefore, we use a repulsive $\Sigma$ well depth of
$U_{\Sigma}^N(n_0)=30$ MeV \cite{Fri94} in this calculation.

The $\sigma^*$-Y coupling constants are obtained by fitting them to a well
depth, ${U_{Y}^{(Y^{'})}}{(n_0)}$, for a hyperon (Y) in a hyperon ($Y^{'}$)
matter at nuclear saturation density \cite{Sch,Sch94}. It is given as
\begin{equation}
U_{\Xi}^{(\Xi)}(n_0) = U_{\Lambda}^{(\Xi)}(n_0) = 2 U_{\Xi}^{(\Lambda)}(n_0)
= 2 U_{\Lambda}^{(\Lambda)}(n_0) = -40
\end{equation}
Nucleons do not couple to the strange mesons i.e.  
$g_{{\sigma^*} N}=g_{\phi N}=0$.

Now we determine kaon-meson coupling constants. According to
the quark model and isospin counting rule, the vector coupling constants are
given by
\begin{equation}
g_{\omega K} = \frac{1}{3} g_{\omega N} ~~~~~ {\rm and} ~~~~~
g_{\rho K} = g_{\rho N} ~.
\end{equation}
The scalar coupling constant is obtained from the real part of
$K^-$ optical potential depth at normal nuclear matter density
\begin{equation}
U_{\bar K} \left(n_0\right) = - g_{\sigma K}\sigma - g_{\omega K}\omega_0 ~.
\end{equation}

The strange meson fields also couple with (anti)kaons. Following
Ref.[\cite{Sch}], the $\sigma^*$-K coupling constant is determined from the
decay of $f_0$(925) as $g_{\sigma^*K}=2.65$, whereas the vector $\phi$ meson
coupling with (anti)kaons is obtained from SU(3) relation as
$\sqrt{2} g_{\phi K} = 6.04$.

It was observed in previous calculations that antikaons experienced an 
attractive potential and kaons had a repulsive interaction in nuclear matter 
\cite{Fri94,Fri99,Koc,Waa,Li,Pal2}. The analysis of $K^-$ atomic data in 
the hybrid model \cite{Fri99} yielded the real part of the antikaon 
optical potential  as large as $U_{\bar K}=-180\pm20$MeV at normal nuclear 
matter density but it was repulsive at low density in accordance with the low 
density theorem. Also, the antikaon potential depth in the coupled channel 
calculation \cite{Koc} was found to be $U_{\bar K}=-100$ MeV whereas the 
chirally motivated coupled channel approach \cite{Waa} 
gave rise to a potential depth of $U_{\bar K}=-120$ MeV.
The different treatments of $\Lambda$(1405) resonance which is considered to be 
an unstable $\bar KN$ bound state just below $K^-p$ threshold, may be 
responsible for the wide range of antikaon optical potential depth values in 
various calculations. In Table II, we list kaon-$\sigma$ meson coupling 
constant $g_{\sigma K}$, for a set of antikaon optical potential depths 
starting from -100 MeV to -180 MeV.

\section{Results and Discussion}
The conversion of nucleons to hyperons is energetically favorable in $\beta$-
equilibrated and charge neutral dense matter. Hyperons first appear in 
dense matter around (2-3)$n_0$. Negatively charged hyperons quench the
growth of electron chemical potential. Also the equation of state is softened
in the presence of hyperons. Therefore, it was shown in various calculations 
neglecting hyperon-hyperon interaction that the formation of $\bar K$
condensate was postponed to higher densities \cite{Ell,Kno,Pal,Bani}. As the 
matter is hyperon-rich
in the high density regime, hyperon-hyperon interaction becomes important. This 
interaction may be accounted by including two additional mesons $\sigma^*$ and 
$\phi$. These strange mesons also couple with (anti)kaons. It follows from 
strangeness numbers of hyperons and $\bar K$ mesons that the 
$\phi$ field is repulsive for hyperons as well as for antikaons. The 
additional attraction due to  $\sigma^*$ field makes the EoS softer. On the
other hand, the repulsive contribution of $\phi$ field becomes dominant at 
higher densities and a stiffer EoS is obtained. For calculations with 
GM1 set and various values of $|U_{\bar K}(n_0)| < 160$ MeV, we find 
that the threshold of $K^-$ condensation in $\beta$-equilibrated hyperonic 
matter is shifted to very high density (7.5$n_0$) where effective 
nucleon mass becomes small \cite{Pal,Bani}. This is attributed to the early 
appearance of $\Xi^-$ hyperons which diminish the electron chemical 
potential and the presence
of repulsive $\phi$ field in the expression of energy for $K^-$ and $\bar K^0$
mesons (Eq.(9)). 

Now we perform our calculation using GM1 set and antikaon 
optical potential depth of $U_{\bar K}(n_0) = -160$ MeV. In 
Figure 1 we exhibit the particle fraction of $\beta$-equilibrated matter 
containing baryons, electrons, muons and $K^-$ mesons. In the pure hadronic 
phase where local charge neutrality condition is imposed, abundances of 
nucleons, electrons and muons increase with density. Here, charge neutrality is
maintained among protons, electrons and muons. With the onset of $K^-$ 
condensation, the mixed phase begins at 2.23$n_0$. This mixed phase is 
determined by the Gibbs phase rules for pure hadronic and  antikaon 
condensed phase
in thermodynamic equilibrium and by global baryon number and electric charge 
conservation laws. We find that $\Lambda$ hyperon is the first strange baryon 
to appear in the mixed phase at 2.51$n_0$. The total baryon density in the
mixed phase is the sum of two contributions from the hadronic and antikaon 
condensed
phase weighted with appropriate volume fractions. As soon as $K^-$ condensate 
is formed, it rapidly grows with density and replaces electrons and muons.
Being bosons, $K^-$ mesons in the lowest energy state are energetically
more favorable to maintain charge neutrality than any other negatively charged 
particles. Consequently, the proton density becomes equal to the density of 
$K^-$ condensate. Also, the density of $\Lambda$ hyperon increases with density 
in the mixed phase. On the other hand, the neutron density decreases in the 
mixed phase. The reason behind it may be the creation of more protons in 
the presence of $K^-$ condensate and also the growth of $\Lambda$ hyperons at 
the expense of neutrons. The mixed phase terminates at 4.0$n_0$. Heavier 
hyperons appear in the pure condensed phase. As $\Xi^-$ hyperons start 
populating this phase at 6.77$n_0$, the density of $K^-$ condensate falls. 
However, the rising behavior of $K^-$ condensate is regained with the onset 
of $\Sigma^+$ hyperons at 7.86$n_0$. It is to be noted that $\Sigma^-$ 
hyperons do not appear in this case.

Besides $K^-$ condensation, we now investigate the role of $\bar K^0$ 
condensation in $\beta$-equilibrated and charge neutral matter consisting of 
nucleons, hyperons, electrons and muons. The abundances of various species are 
shown in Figure 2. The composition of the pure 
hadronic and mixed phase in this case  is similar to that of Fig. 1. Once
the mixed phase is over, $\bar K^0$ condensation occurs at $\sim$4.1$ n_0$. 
We consider  $\bar K^0$ condensation as a second order phase transition. 
The antikaon condensed phase including $\bar K^0$ condensate along with 
$K^-$ condensate is different from that of Fig. 1. With the appearance 
of $\bar K^0$ condensate, neutron and proton abundances become equal and a
symmetric matter of nucleons is formed in this phase. The density of 
$\bar K^0$ condensate increases with baryon density uninterruptedly and even 
becomes larger than the density of  $K^-$ condensate. 
At higher densities, more hyperons start populating the system. As soon as 
negatively charged hyperons - $\Xi^-$ and $\Sigma^-$ appear, the density of  
$K^-$ condensate is observed to fall drastically. 
This is quite expected because it is energetically favorable for particles 
carrying  conserved baryon numbers to achieve charge neutrality in the system.
Leptons or mesons are no longer required for this sole purpose. Moreover, 
lepton number or meson number is not conserved in the star. 
The system is dominated by $\bar K^0$ condensate in the high density regime.

One important aspect of this calculation is the dynamic behavior of baryons in 
pure hadronic and antikaon condensed phase. 
Earlier, this dynamical behavior of nucleons was noted by Glendenning 
and Schaffner \cite{Gle99}. Besides nucleons, we find similar properties of 
$\Lambda$ hyperons which appear in the mixed phase. It is evident
from  Eq. (11) and Eq. (12) that baryons have different effective masses in the 
hadronic and condensed phase due to different $\sigma$ and $\sigma^*$ fields. 
In the pure hadronic phase, the effective mass of a baryon decreases with 
density. With the onset of the mixed phase, a second solution for the effective 
baryon mass emerges. This is the effective baryon mass 
in the condensed phase. The value of the effective baryon mass 
in the pure hadronic phase is always higher than 
that of the antikaon condensed phase. In the condensed phase, the effective 
mass of a baryon also decreases with density.

Equation of state, pressure(P) versus energy density  ($\varepsilon$), for 
$\beta$-equilibrated and charge neutral matter with different compositions are 
displayed in Figure 3. Here the long-dashed curve stands for hyperonic matter 
without any antikaon condensate. The dotted curve represents hyperonic matter 
including $K^-$ condensate whereas the solid curve corresponds to hyperonic 
matter with both $K^-$ and $\bar K^0$ condensate. The EoS denoted by the 
dashed-dot curve is same as given by the solid except no $\Xi$ hyperon is 
present in the system in this case. Two kinks in the solid curve marks 
the beginning and end of the mixed phase where pure hadronic and $K^-$ 
condensed phase are in thermodynamic equilibrium as dictated by the Gibbs 
conditions and global conservation laws. Immediately after the termination of 
the mixed phase at 4.0$n_0$, $\bar K^0$ condensation sets in at 
$\sim$ 4.1$n_0$. 
We have treated here $\bar K^0$ condensation  as a second order phase 
transition. Also we find that the pressure increases continuously with energy 
(solid curve) after the formation of $\bar K^0$ condensation. Earlier
it was noted that the  antikaon condensation could be a second order phase 
transition depending on the strength of antikaon optical potential depth and 
parameter sets of different models \cite{Gle99,Pal,Bani}. Though the dotted 
curve has overlap with the solid curve up to the mixed phase, it becomes stiffer
in the high density regime. On the other hand, the dashed-dot curve 
becomes stiffer compared to the solid curve because there is lesser number 
of degrees of freedom with the absence of $\Xi$ hyperons in the former case. 
With the appearance of both $K^-$ and $\bar K^0$ condensation in hyperonic 
matter, the overall EoS (solid curve) is softened due to the strong attraction 
imparted by antikaon condensation.  

We have used the results of Baym, Pethick and Sutherland \cite{Bay} to describe
the crust of a compact star
composed  of leptons and nuclei for the low density ($n_B < 0.001 fm^{-3}$)
EoS. In the mid density regime ($0.001 \leq n_B < 0.08 fm^{-3}$)
the results of Negele and Vautherin \cite{Neg} are taken into account. Above 
this density, an EoS calculated in the relativistic model has been adopted.

Now we present the results of static structures of  spherically symmetric
compact stars calculated using TOV equations. The static
compact star sequences representing the stellar masses $M/M_\odot$ and
the corresponding central energy densities $\varepsilon_c$ are shown in
Figure 4. We denote the sequence of neutron stars with $K^-$ condensation in 
hyperonic matter by the dotted curve. Here, the maximum neutron star mass 
is 1.649$M_\odot$ occurring at central density 7.7$n_0$. The maximum neutron
star mass consisting only of hyperons is 1.789$M_{\odot}$ corresponding to
the central density 5.16$n_0$ (not shown in the Figure). The solid curve 
represents compact stars  with 
both $K^-$ and $\bar K^0$ condensation in hyperonic matter.
When $\bar K^0$ condensation along with $K^-$ condensation is taken 
into account, the limiting mass of the neutron star branch is 
attained at much earlier central density 4.49$n_0$. In fact, the limiting mass 
star appears  very close to the upper boundary of the mixed phase and has a 
smaller value 1.571$M_\odot$ than the situation with only $K^-$  condensation. 
This reduction in the maximum mass of the star may be attributed to the
softening of the EoS in the presence of $K^-$ and  $\bar K$ condensate.
After the stable neutron star branch, there is an unstable region followed 
by a sequence of compact stars in Fig. 4. This branch  of super dense 
stars beyond the neutron star branch is called a third family of compact 
stars \cite{Ger,Gle00,Sche}. There are two third family branches in Fig. 4.
The lower curve depicts the third family of compact stars with both antikaon
condensates in hyperonic matter (case I) whereas the upper curve represents 
the third family of compact stars including both condensates in hyperonic 
matter except $\Xi$ hyperons (case II). The maximum masses of compact stars in 
the third family branches denoted by the lower curve  and upper curve 
are 1.552$M_\odot$ and 1.561$M_\odot$
corresponding to the central densities 8.09$n_0$ and $9.20n_0$ respectively.
It is the behavior of the EoS including both $K^-$ and $\bar K^0$ condensate 
in the mid and high density regime that results in a third family of super 
dense 
star. The possibility of a third family of compact stars was already discussed 
in the context of quark-hadron phase transition \cite{Gle00,Sche} and hadronic 
matter to hyperonic matter phase transition \cite{Han}.

We have found a new sequence of compact stars called the third family of 
compact stars beyond the neutron star branch as is evident in Fig. 4. 
So far, we have discussed the hydrostatic stability
of the stellar sequence in the third family using TOV equations \cite{Gle}. 
In Fig. 4, we note that the neutron star branch or the third family branch has
positive slope i.e. dM/d$\varepsilon_c > 0$. Though the positive slope of a 
steller sequence is a necessary condition for stability, dynamical 
stability requires the analysis                     
of the fundamental mode of radial vibration \cite{Gle00,Sche,Gle95,Bar,Mel}. 
In the third family branches in Fig. 4, we  find that each  configuration is 
stable because the squared frequency of the fundamental mode associated with it 
is positive. Similar results were obtained in the study of compact stars 
beyond the neutron star branch using quark-hadron phase transition 
by various groups \cite{Gle00,Sche}.

We exhibit the mass-radius relationship for different EoS in Figure 5. The 
maximum mass star composed of hyperons and no condensate (not shown here)
has a radius of 12.8 Km whereas it is 10.9 Km for the maximum mass neutron star 
including hyperons and $K^-$ condensate (dotted curve). Two cases where 
we obtain stable solutions  beyond the neutron star branch
are shown in the inset. The solid curve denotes the sequences of compact stars 
having hyperons and $K^-$ and $\bar K^0$ condensate. In the neutron star 
branch, the maximum mass star has a radius of 12.8 Km. For the third 
family branches, the lower curve corresponds to case I and the upper
curve denotes case II. The radius of the maximum mass super dense star on the
lower curve is 10.7 Km whereas that of case II is 10.2 Km. In this calculation, 
we find that the existence of non-identical stars of same mass may be 
possible because of partial overlapping of the neutron star and the third 
family branch. These pairs are known as "neutron star twins" \cite{Gle00,Sche}. 
Also, we observe that the high density twin is a very compact object with a 
different composition and smaller radius than its low density counter part. 
It is to be noted that  the radius of the maximum mass star with $K^-$ 
condensate in hyperonic matter is comparable to that of the third family 
considered here. This implies that only one star with small radius  cannot 
ensure the existence of a third family of compact stars. 
Schertler et al. \cite{Sche} argued that two stars - one in the neutron star 
branch and the other in the third family branch with similar masses 
but with a radius difference of a few kilometer might reveal the existence of 
a third family of compact stars. In this case, the mass difference 
($\Delta $M) of two stars is to be much less than the radius difference 
($\Delta $R) i.e.  $\Delta M << \Delta R$. In our calculation, 
we note that $\Delta $R is almost two orders of magnitude larger than  
$\Delta $M.    

We also investigate the high density behaviour of the equation of state in TM1 
model \cite{Tok}. Besides scalar self interaction terms, TM1 model includes
non-linear $\omega$ meson term \cite{Tok,Bod}. The parameters of TM1 model are
listed in Table I and Table II. In this model, antikaon condensations are 
found to be second order phase transitions for all values 
of antikaon optical potential depth \cite{Gle99,Pal,Bani}. 
The EoS behaves smoothly even in the high density
regime. There is no existence of a third family of compact stars in TM1 model
because of this smooth EoS. 

\section{Summary and conclusions}

We have investigated $K^-$ and $\bar K^0$-meson condensation in 
$\beta$-equilibrated hyperonic matter within a relativistic mean field model. 
In this model, baryon-baryon and (anti)kaon-baryon interactions are mediated by
the exchange of $\sigma$, $\omega$ and $\rho$ meson. Also, hyperon-hyperon 
interaction is accounted by considering two additional strange mesons 
${\sigma} ^*$ and $\phi$. In this calculation, we have adopted GM1 
parameter set. The coupling constants of (anti)kaon-meson are determined from
a quark model and empirically known values of antikaon optical potential depth
at normal nuclear matter density. Similarly, the hyperon-meson coupling 
constants are obtained from hypernuclear data and SU(6) symmetry relations.
However, the coupling constant of hyperons to ${\sigma}^*$ meson is obtained
by reproducing the hyperon depth in a bath of the hyperons at normal nuclear 
matter density. 

The condensation of $K^-$ meson in hyperonic matter is described as a first
order phase transition in this calculation. The $\beta$-equilibrated matter is
composed of three phases - pure hadronic phase, pure $K^-$ condensed phase and
the mixed phase of two pure phases. Along with the Gibbs phase rules, global 
conservation laws are employed in the mixed phase.

The appearance of antikaon condensation is quite sensitive to the EoS and also
depends on the value of antikaon optical potential depth. In the presence of 
hyperons, an EoS becomes softer thus delaying the onset of antikaon 
condensation. For antikaon optical potential depth $|U_{\bar K}(n_0)| < 
160$ MeV, the early appearance of negatively 
charged hyperons in particular ${\Xi}^-$ hyperons,
reduces the electron chemical potential. And the antikaon condensation does not
occur even at 7.5$n_0$. In our calculation with $U_{\bar K}(n_0) = -160$ MeV,
the first order phase transition to $K^-$ condensation begins at 4.0$n_0$.
The pure hadronic phase does not contain any hyperon whereas only $\Lambda$
hyperons appear in the mixed phase. As soon as $K^-$ condensation sets in, it 
is energetically favourable for $K^-$ mesons in the zero momentum state to 
maintain charge neutrality in the system. Consequently, the formation of
negatively charged 
hyperons are delayed to higher density. Immediately after the end of the mixed
phase, $\bar K^0$ condensation occurs in the pure $K^-$ condensed phase. Here,
we have treated $\bar K^0$ condensation as a second order phase transition. With
the formation of $\bar K^0$ condensation, abundances of neutron and protons
become equal. Negatively charged hyperons - ${\Xi}^-$ and $\Sigma^-$ start
populating the system around $\sim 7n_0$. And the density of $K^-$ condensate
dwindles because hyperons having conserved baryon number are 
energetically more favourable for charge neutrality. Consequently, $\bar K^0$
condensate dominates the high density regime.

The equation of state becomes softer in the presence of antikaon condensate 
compared with the situation without any condensate. Among all EoS considered
here, the softest EoS is the one that includes both $K^-$ and $\bar K^0$ 
condensate. This softening leads to the reduction in maximum masses of compact 
stars. The behaviour of the EoS with both antikaon condensates at very high
density $\sim 8-10 n_0$ and its connection to the possible existence of
stable super dense stars beyond the neutron star branch have been investigated
in this work. With the EoS corresponding to hyperonic matter with $K^-$ and
$\bar K^0$ condensate, we find, after the termination of the neutron star
branch, an unstable region followed by a stable sequence of compact stars
called the third family of compact stars. It is noted that neutron star twins
could exist because of partially overlapping mass regions of the neutron star
branch and the third family branch. The high density twin in the third family 
branch has a different composition and smaller radius than its low density 
counter part. The observation of two compact stars with almost similar masses
but different radii may shed light on the existence of a third family of super
dense stars.   


\begin{table}

\caption{The nucleon-meson coupling constants in 
the GM1 set are taken from Ref. [24]. In this relativistic model, the baryons
interact via nonlinear $\sigma$-meson and linear $\omega$-meson exchanges. 
The coupling constants are obtained by reproducing the nuclear matter 
properties of binding energy $E/B=-16.3$ MeV, baryon density $n_0=0.153$ 
fm$^{-3}$, asymmetry energy coefficient $a_{\rm asy}=32.5$ MeV, 
incompressibility $K=300$ MeV, and effective nucleon mass $m^*_N/m_N = 0.70$. 
The hadronic masses are $m_N=938$ MeV, $m_\sigma=550$ MeV, 
$m_\omega=783$ MeV, and $m_\rho=770$ MeV. The parameter set
TM1 is obtained from Ref. [42] which incorporates nonlinear exchanges in 
both $\sigma$ and $\omega$ mesons. The nuclear matter properties in the TM1 set
are $E/B=-16.3$ MeV, $n_0=0.145$ fm$^{-3}$, $a_{\rm asy}=36.9$ MeV, 
$K=281$ MeV, and $m^*_N/m_N = 0.634$. All the hadronic masses in 
this model are same as GM1 except for $\sigma$-meson which is 
$m_\sigma=511.198$ MeV. All the parameters are dimensionless, except $g_2$ 
which is in fm$^{-1}$.}

\begin{tabular}{ccccccc} 

\hfil& $g_{\sigma N}$& $g_{\omega N}$& $g_{\rho N}$&
$g_2$& $g_3$& $g_4$ \\ \hline
GM1& 9.5708& 10.5964& 8.1957& 12.2817& -8.9780& $-$ \\
TM1& 10.0289& 12.6139& 4.6322& -7.2325& 0.6183& 71.3075 \\

\end{tabular}
\end{table}

\begin{table}

\caption{The coupling constants for antikaons ($\bar K$) to
$\sigma$-meson, $g_{\sigma K}$, for various values of $\bar K$ optical 
potential depths $U_{\bar K}$ (in MeV) at the saturation density. The 
results are for the GM1 and TM1 set.}

\begin{tabular}{cccccc} 

$U_{\bar K}$& -100& -120& -140& -160& -180 \\ \hline
GM1& 0.9542& 1.6337& 2.3142& 2.9937& 3.6742\\
TM1& 0.2537& 0.8384& 1.4241& 2.0098& 2.5955 \\

\end{tabular}
\end{table}
\newpage 
\vspace{-2cm}

{\centerline{
\epsfxsize=12cm
\epsfysize=14cm
\epsffile{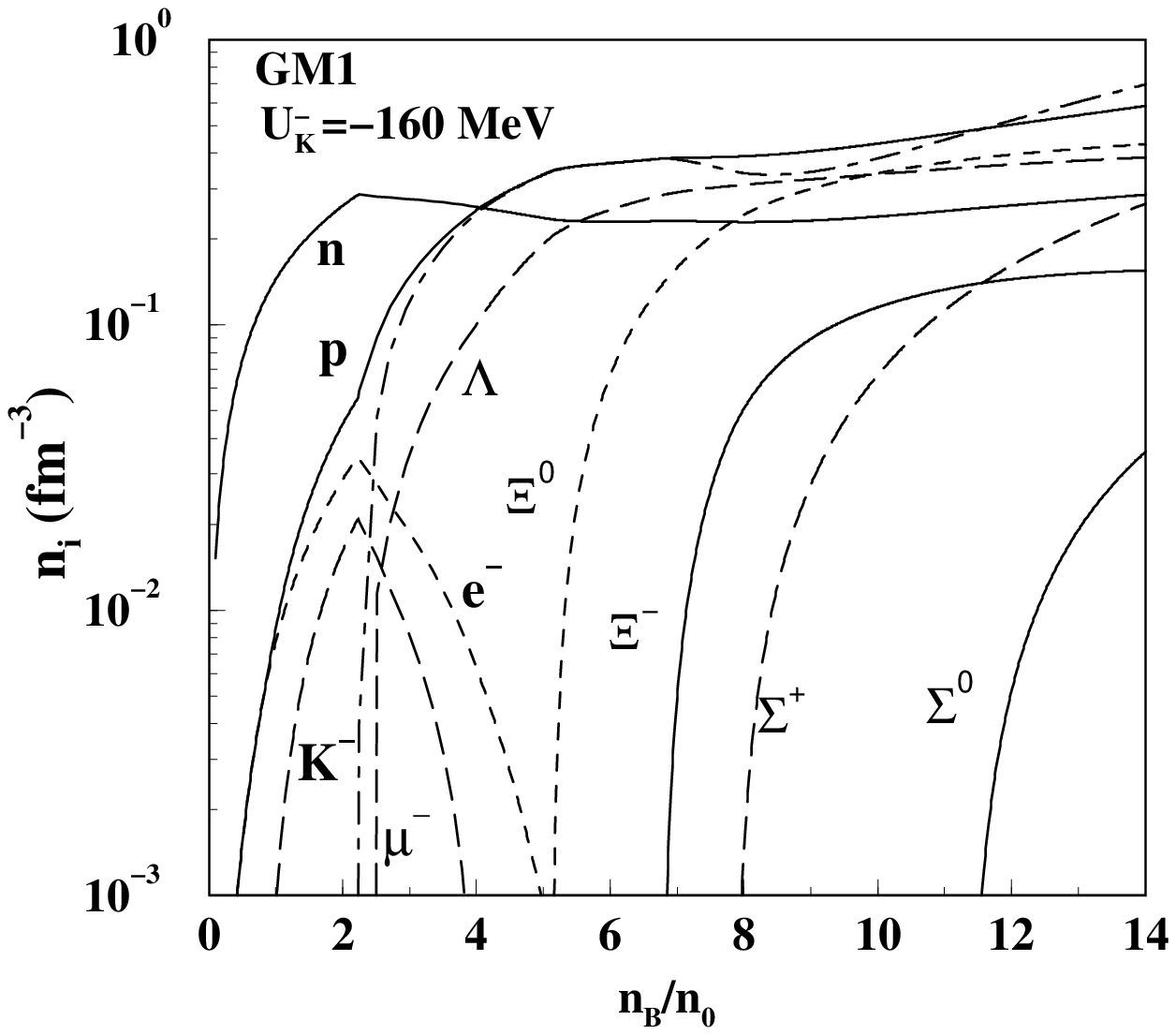}
}}

\vspace{4.0cm}

\noindent{\small{
FIG. 1. The proper number densities $n_i$ of various compositions in 
$\beta$-equilibrated hyperonic matter including $K^-$ condensate for GM1 
model and  antikaon optical potential depth at normal nuclear matter density 
$U_{\bar K} = -160$ MeV.}}

\newpage
\vspace{-2cm}

{\centerline{
\epsfxsize=12cm
\epsfysize=14cm
\epsffile{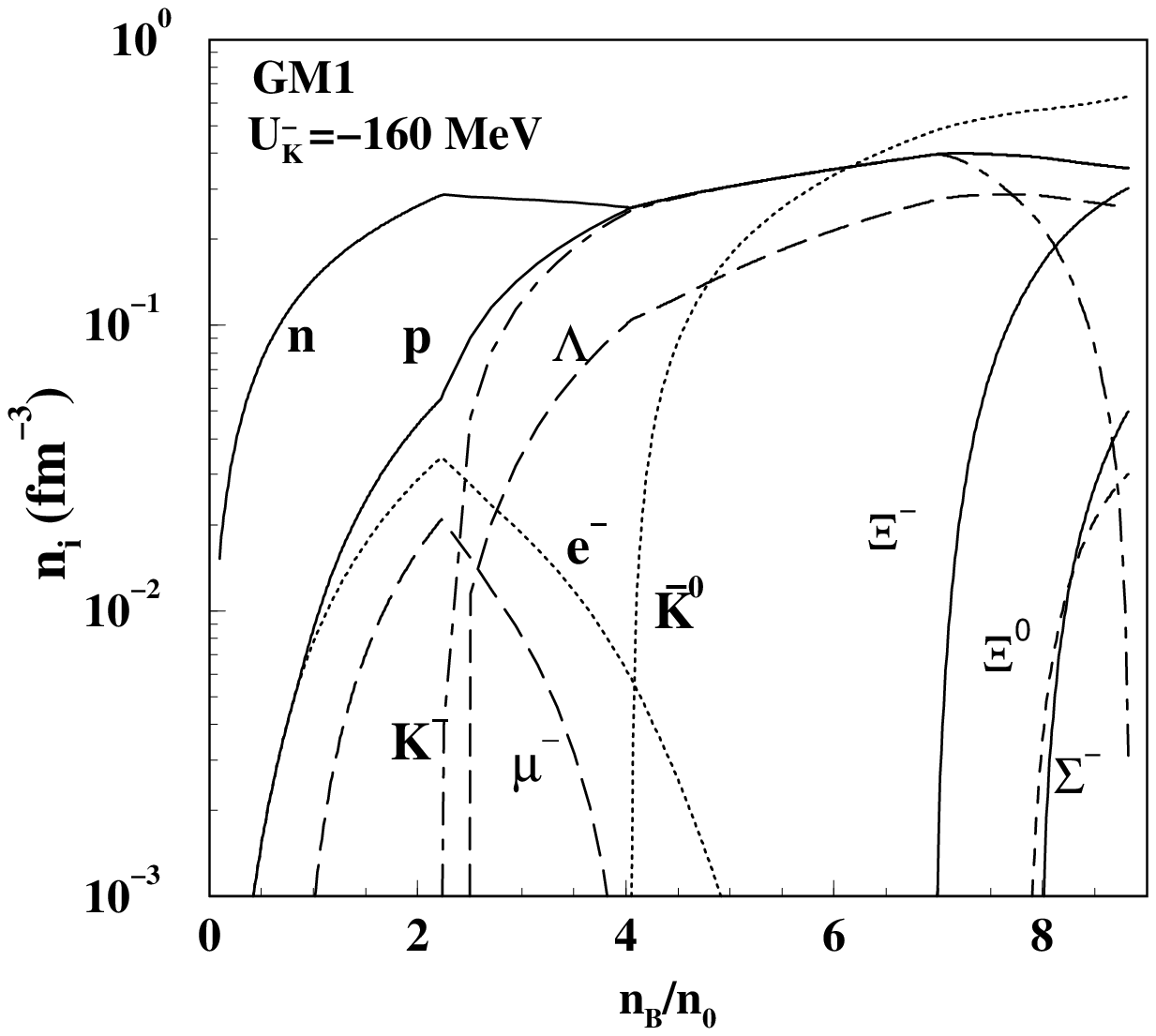}
}}

\vspace{4.0cm}

\noindent{\small{
FIG. 2. The proper number densities $n_i$ of various compositions in 
$\beta$-equilibrated hyperonic matter including both $K^-$ and $\bar K^0$
condensate for GM1 model and  antikaon optical potential depth at normal 
nuclear matter density $U_{\bar K} = -160$ MeV.}}
 \newpage
\vspace{-2cm}

{\centerline{
\epsfxsize=20cm
\epsfysize=22cm
\epsffile{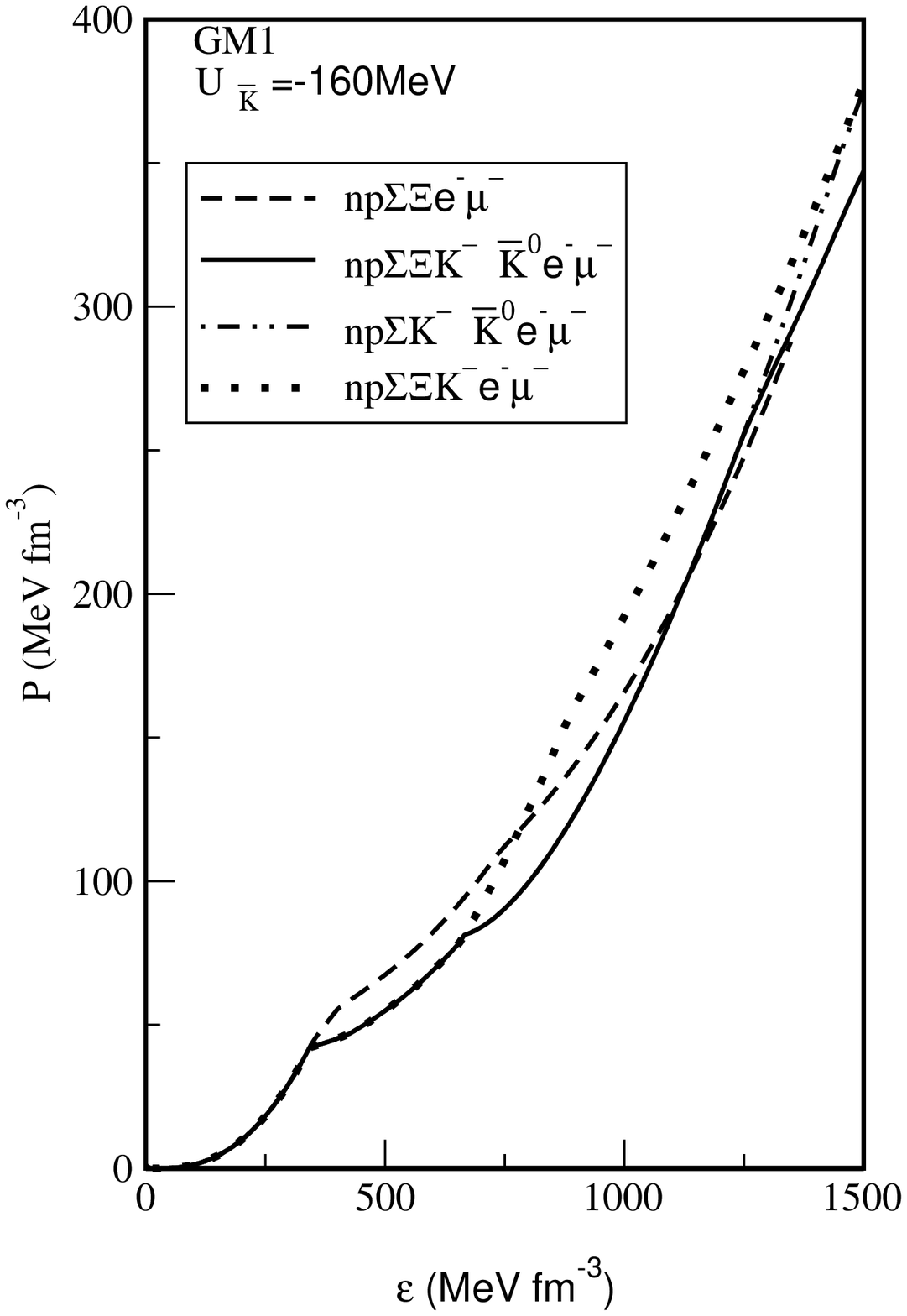}
}}

\vspace{-1.5cm}

\noindent{\small{
FIG. 3. The equation of state, pressure $P$ vs. energy density $\varepsilon$ 
in GM1 model. The results are for  hyperonic matter(dashed line), hyperonic
matter including $K^-$ condensate (dotted line) and $K^-$ and $\bar K^0$ 
condensation in hyperonic matter (solid line) and in hyperonic matter excluding
$\Xi$ (dashed-dot line) calculated with the antikaon optical potential 
depth at normal nuclear matter density of $U_{\bar K}= -160$ MeV. }}
 \newpage

{\centerline{
\epsfxsize=18cm
\epsfysize=22cm
\epsffile{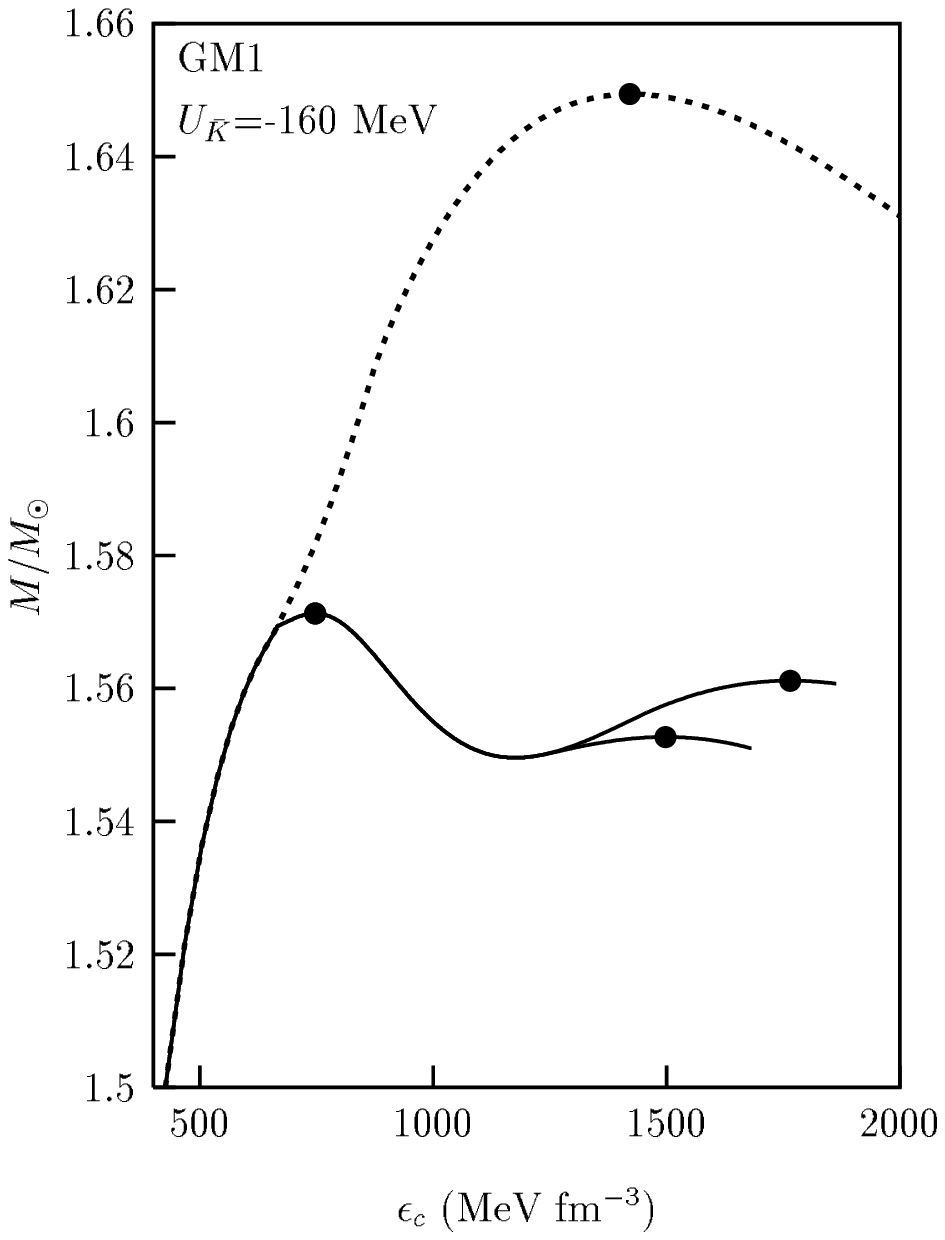}
}}

\vspace{-5.0cm}

\noindent{\small{
FIG. 4. The compact star mass sequences are plotted with central energy 
density for GM1 model and the antikaon optical potential depth 
of  $U_{\bar K}=-160$ MeV. The star masses of hyperonic matter with $K^-$
condensate and with further inclusion of $\bar K^0$ condensate are shown  
here. In the latter case, a second sequence of compact stars appears in the
high energy density regime. }}
 \newpage
\vspace{-2.0cm}
{\centerline{
\epsfxsize=18cm
\epsfysize=22cm
\epsffile{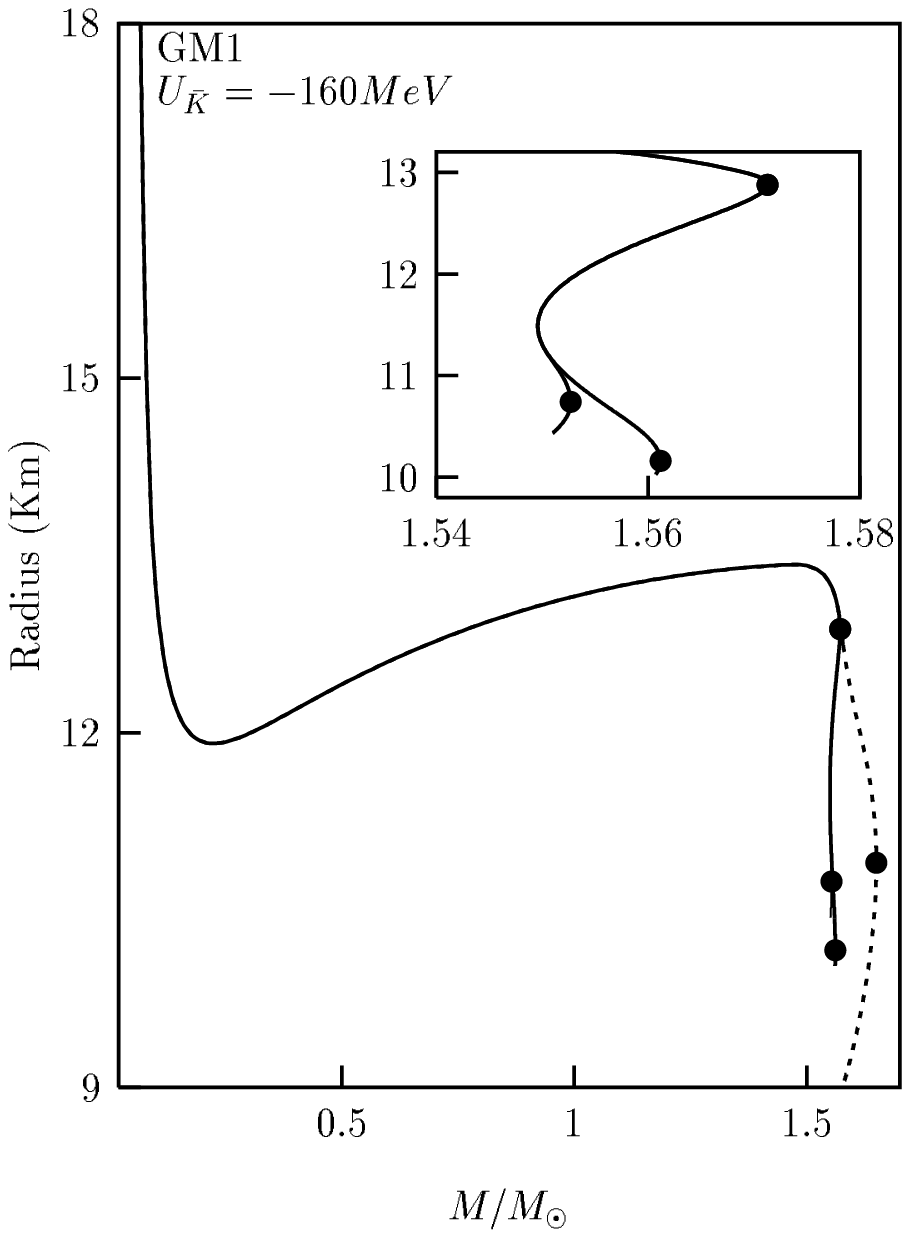}
}}
\vspace{-5.0cm}
\noindent{\small{
FIG. 5. The mass-radius relationship for compact star sequences for hyperonic
matter with $K^-$ condensate and with further inclusion of $\bar K^0$ 
condensate for GM1 model and antikaon optical potential depth of 
$U_{\bar K} = - 160$ MeV. The mass-radius relationship for the third family 
branch is shown in the inset.}}

\end{document}